\documentclass[
showkeys,12pt,
preprint,preprintnumbers,nofootinbib,
groupedaddress,superscriptaddress,amsmath,amssymb]{revtex4}
\usepackage[dvipdfmx]{graphicx}
\usepackage{dcolumn}
\usepackage{bm}
\usepackage{amssymb}
\usepackage{amsmath}
\usepackage{epsfig}    
\usepackage[dvipdfmx]{color}
\usepackage{slashed}
\usepackage{hhline}
\usepackage{soul}

\def\be{\begin{equation}}
\def\ee{\end{equation}}
\newcommand{\bea}{\begin{eqnarray}}
\newcommand{\eea}{\end{eqnarray}}
\newcommand{\nn}{\nonumber}

\newcommand{\ET}{\mbox{$\not \hspace{-0.10cm} E_T$ }}

\numberwithin{equation}{section}

\begin{document}

\title{An explanation of one loop induced $h\to \mu \tau$ decay  }
\preprint{KIAS-P16030}
\author{Seungwon Baek}
\email{swbaek@kias.re.kr}
\affiliation{School of Physics, KIAS, Seoul 130-722, Korea}
\author{Takaaki Nomura}
\email{nomura@kias.re.kr}
\affiliation{School of Physics, KIAS, Seoul 130-722, Korea}
\author{Hiroshi Okada}
\email{macokada3hiroshi@gmail.com}
\affiliation{Physics Division, National Center for Theoretical Sciences, Hsinchu, Taiwan 300}

\date{\today}

\begin{abstract}
We discuss a possibility to explain the excess of  $h\to\mu\tau$ at one-loop level. We introduce three generation of vector-like lepton
doublet $L'$ and two singlet scalars $S_{1,2}$ which are odd under  $Z_2$, while all the standard model fields are even under
this discrete symmetry. We show that $S_1$ can be a good dark matter candidate.
We show that we can explain the dark matter relic abundance, large part of the discrepancy of muon $g-2$ between experiments and 
the standard model predictions, as well as the $h\to\mu\tau$ excess of $\sim 1$ \%, while evading constraints from
experiments of dark matter direct detection and charged lepton flavor violating processes.
We also consider prospects of production of $S_2$ at LHC with energy $\sqrt{s}=14$ TeV.
\end{abstract}
\maketitle 
\newpage

\section{Introduction}

The mechanism of the electroweak symmetry breaking is being further understood by the discovery of the Standard Model(SM)-like Higgs boson with mass around 125 GeV at the ATLAS~\cite{Chatrchyan:2012xdj} and CMS~\cite{Aad:2012tfa} experiments.
Furthermore, we would obtain some insight on physics beyond the SM by exploring nature of the Higgs boson such as its decay channels and scalar potential.

The CMS~\cite{Khachatryan:2015kon} and ATLAS~\cite{Aad:2015gha} collaborations reported the results on their search for rare Higgs
decay $h \to \mu \tau$ with the dataset obtained at the LHC 8 TeV.
An excess of the events was observed by CMS, with a significance of 2.4$\sigma$, where the best fit value of branching ratio is $BR(h \to \mu \tau) = (0.84^{+0.39}_{-0.37})\%$.
ATLAS'  best fit value is $BR(h \to \mu \tau) = (0.77\pm 0.62)\%$,  consistent with but less significant than CMS.
Since the lepton flavor violating Higgs decay is highly suppressed in the SM, their findings are an intriguing hint indicating new
physics (NP) which induces lepton flavor violation in the charged lepton sector, although we need more data to get conclusive
evidence for NP.
Actually, inspired by the excess, new physics effects in $h \to \mu \tau$ decay have been studied in~\cite{Campos:2014zaa,Sierra:2014nqa,Lee:2014rba, Heeck:2014qea,Crivellin:2015mga,Dorsner:2015mja,Omura:2015nja,Crivellin:2015lwa,Das:2015zwa,Bishara:2015cha,Varzielas:2015joa,He:2015rqa,Chiang:2015cba,Altmannshofer:2015esa,Cheung:2015yga,Arganda:2015naa,Botella:2015hoa,Baek:2015mea,Huang:2015vpt,Baek:2015fma,Arganda:2015uca,Aloni:2015wvn,Benbrik:2015evd,Omura:2015xcg,Zhang:2015csm,Hue:2015fbb,Bizot:2015qqo,Han:2016bvl,Chang:2016ave,Chen:2016lsr,Alvarado:2016par,Banerjee:2016foh,Hayreter:2016kyv,Huitu:2016pwk,Chakraborty:2016gff,Lami:2016mjf,Thuc:2016qva}.
Earlier works on the flavor violating Higgs decay can be found in~\cite{Arganda:2004bz, Blankenburg:2012ex,Arhrib:2012mg,Harnik:2012pb,Dery:2013rta,Arana-Catania:2013xma,Arroyo:2013tna,Celis:2013xja,Falkowski:2013jya,Arganda:2014dta,Dery:2014kxa}.

In this paper, we investigate lepton flavor violating effect which is mediated by an exotic lepton doublet $L'$ and inert singlet scalars $S$ which are odd under discrete symmetry $Z_2$.
The interaction term $\bar L L' S$ allowed by both the SM gauge and the $Z_2$ symmetries often appears in radiative
seesaw models, providing active neutrino masses. The neutral components of $L'$ or $S$ can be also good dark matter(DM) candidates.
In addition lepton flavor violating (LFV) Higgs decay can be induced at one-loop level with the interaction. 
Thus this interaction provides interesting effects connecting active neutrino masses, dark matter, and lepton flavor violating Higgs decays.
Focusing on the interaction, we explore $h \to \mu \tau$, charged lepton flavor violations, anomalous magnetic moment of muon and relic density of bosonic dark matter candidate, considering a specific model as an example. 
Then we search for the parameter region which explains the excess of $h \to \mu \tau$ observed by CMS with sizable muon magnetic moment and observed relic density of DM, 
taking into account the constraints from flavor violating lepton decays.
Furthermore we discuss possible signature of our scenario which could be tested at the LHC.



This paper is organized as follows.
In Sec.~II, we show our model,  including LFVs, muon anomalous magnetic moment, and  LFV Higgs decay.
In Sec.~III, we carry out numerical analysis including bosonic DM candidate to explain relic density and direct detection in a specific case.
We conclude and discuss in Sec.~IV.


\section{ Model setup}
 \begin{widetext}
\begin{center} 
\begin{table}
\begin{tabular}{|c||c|c|c||c|c|}\hline\hline  
&\multicolumn{3}{c||}{Lepton Fields} & \multicolumn{2}{c|}{Scalar Fields} \\\hline
& ~$L_L$~ & ~$e_R^{}$~ & ~$L'_{L(R)}$ ~ & ~$\Phi$~ & ~$S^{m}$ \\\hline 
$SU(2)_L$ & $\bm{2}$  & $\bm{1}$  & $\bm{2}$ & $\bm{2}$ & $\bm{1}$ \\\hline 
$U(1)_Y$ & $-\frac12$ & $-1$  & $-\frac{N}{2}$ & $\frac12$  & $\frac{-1+N}{2}$  \\\hline
 $Z_2$ & $+$ & $+$   & $-$ & $+$  & $-$  \\\hline
\end{tabular}
\caption{Contents of fermion and scalar fields
and their charge assignments under $SU(2)_L\times U(1)_Y$, where $m\equiv \frac{-1+N}{2}$ is the quantum number of the electric charge.}
\label{tab:1}
\end{table}
\end{center}
\end{widetext}

In this section, we explain our model. 
The particle contents and their charges are shown in Table~\ref{tab:1}.
We add {three} iso-spin doublet vector-like exotic fermions $L'$ with hypercharge $-N/2$,
 and an isospin singlet scalar $S^m$  with $(-1+N)/2$ hypercharge to the SM, where they are odd under $Z_2$, $N (\ge 1)$ is an odd integer and
 $m(\equiv \frac{N-1}{2})$ is 
the electric charge of $S$. 
Then we define the exotic lepton as 
\begin{align}
L'\equiv [\Psi^{-m},\Psi^{-m-1}]^T.
\end{align}
We assume that  only the SM Higgs $\Phi$ have vacuum
expectation value (VEV), which is symbolized by $v/\sqrt2$. 

The relevant Lagrangian and Higgs potential under these symmetries are given by
\begin{align}
-\mathcal{L}_{Y}
&=
(y_{\ell})_{ij} \bar L_{Li} \Phi e_{Rj}  + (y_L)_{ij} \bar L_{L_i} L'_{R_j} S^m
 + { (M_{L})_{ij}} \bar L'_{Li} L'_{Rj} + {\rm h.c.}, \nn\\
{\cal V}&=
m^2_{\Phi} \Phi^\dag\Phi + m^2_{S} |S^m|^2 
+ \lambda_{\Phi} |\Phi^\dag\Phi|^2  + \lambda_S |S^m|^4
+ \lambda_{\Phi S} |\Phi|^2 |S^m|^2 
\label{Eq:lag-flavor}
\end{align}
where  the first term of $\mathcal{L}_{Y}$ can generates the SM
charged-lepton masses $m_\ell\equiv y_\ell v/\sqrt2$ after the spontaneous electroweak symmetry breaking by VEV of $\Phi$.
{We assume all the coefficients are real and positive for simplicity. }
The scalar fields can be parameterized as 
\begin{align}
&\Phi =\left[
\begin{array}{c}
w^+\\
\frac{v+h+iz}{\sqrt2}
\end{array}\right],
\label{component}
\end{align}
where $v~\simeq 246$ GeV is VEV of the Higgs doublet, and $w^\pm$
and $z$ are Goldstone bosons 
which are absorbed by the longitudinal component of $W$ and $Z$ boson, respectively.
Inserting the tadpole condition; $\partial\mathcal{V}/\partial\phi|_{v}=0$,
the SM Higgs mass is given by $\sqrt{2\lambda_\Phi}v$. The mass eigenstate $S^{\prime m}$ of the exotic scalar has mass
\begin{align}
& m_{S^{\prime}}= m_S^2+\frac{\lambda_{\Phi S}v^2}{2}.
\end{align}



\subsection{ Lepton Flavor Violations and Muon anomalous magnetic moment}\label{lfv-lu}
\begin{table}[t]
\begin{tabular}{c|c|c} \hline
Process & $(b,a)$ & Experimental bounds ($90\%$ CL) \\ \hline
$\mu^{-} \to e^{-} \gamma$ & $(2,1)$ &
	$\text{Br}(\mu \to e\gamma) < 5.7 \times 10^{-13}$  \\
$\tau^{-} \to e^{-} \gamma$ & $(3,1)$ &
	$\text{Br}(\tau \to e\gamma) < 3.3 \times 10^{-8}$ \\
$\tau^{-} \to \mu^{-} \gamma$ & $(3,2)$ &
	$\text{Br}(\tau \to \mu\gamma) < 4.4 \times 10^{-8}$  \\ \hline
\end{tabular}
\caption{Summary of $\ell_b \to \ell_a \gamma$ process and the lower bound of experimental data~\cite{Adam:2013mnn}.}
\label{tab:Cif}
\end{table}

First, let us consider the LFV decays in the charged lepton sector, which impose constraints on the $h\to\mu\tau$ anomaly.
They are summarized in Table~\ref{tab:Cif}.
The processes $\ell_b\to\ell_a \gamma (b>a)$ arise from one-loop diagrams through the  term $(y_L)_{ij}  (\bar\ell_L)_i (\Psi^{-m-1})_jS^m$.
Then their branching ratios ${\rm BR}(\ell_b\to\ell_a \gamma)$ are defined by
\begin{align}
{\rm BR}(\ell_b\to\ell_a \gamma)
=
\frac{48\pi^3 \alpha_{\rm em}C_b}{{\rm G_F^2} m_b^2 }(|(a_R)_{ab}|^2+|(a_L)_{ab}|^2),
\end{align}
where $\alpha_{\rm em}$ is the fine structure constant, $C_b \approx (1,1/5)$ for ($b=\mu,\tau$), ${\rm G_F}\approx1.17\times 10^{-5}$ GeV$^{-2}$ is the Fermi constant. 
The Wilson coefficients $(a_R)_{ab}$ are obtained to be
\begin{align}
(a_R)_{ab}&=-\frac{m_b}{(4\pi)^2} 
\sum_{i=1}^3
(y^\dag_L)_{ai} (y_L)_{ib} \left[ (m+1)F[m_{S^m},M_{\Psi^{m+1}}]+m F[M_{\Psi^{m+1}},m_{S^m}] \right],\\
F[m_a,m_b]&\equiv
\frac{2 m_a^6 + 3 m_a^4 m_b^2 + 12 m_a^4 m_b^2 \ln\left[\frac{m_b}{m_a}\right]-6 m_a^2 m_b^4+m_b^6}{12(m_a^2-m_b^2)^4},
\label{eq:lfvs}
\end{align} 
while the chirality-flipped ones are suppressed by small mass ratios: $a_L=a_R {m_{a}}/{m_{b}}$.
Here $M_{\Psi^{1+m}}(=M_{\Psi^m})=M_L$.

Our formula of the muon anomalous magnetic moment (muon $g-2$) is also given in terms of $a_{L/R}$ by
\begin{align}
\Delta a_\mu\approx -{m_\mu}(a_R)_{22},\label{eq:damu}
\end{align}
where the lower index $2$ of $a_{R}$ is muon eigenstate.

\subsection{$h\to \mu\tau$ excess}
\begin{figure}[tb]
\begin{center}
\includegraphics[scale=0.6]{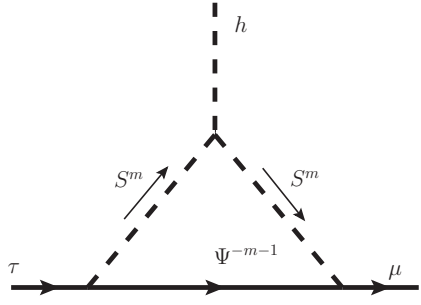}
\caption{One-loop Feynman diagram for $h \to \mu \tau$. The Higgs ($h$) line can also be attached to external $\mu$ or
  $\tau$ lines.}
\label{fig:feynman}
\end{center}
\end{figure}

In our case, the excess of $h\to \mu\tau$ can be generated at one-loop level as the leading contribution. Its Feynman 
diagram is shown in Fig.~\ref{fig:feynman}.

The resultant decay rate formulas are expressed as
\begin{align}
& \Gamma(h\to \mu\tau) = \frac{|\bar M|^2}{8\pi m_h^2}
\sqrt{\frac{(m_h+m_\mu)^2-m_\tau^2}{2m_h} \frac{(m_h-m_\mu)^2-m_\tau^2}{2m_h} } ,\\
&|\bar M|^2=
\sum_{i=1}^3
\frac{|(y^\dag_L)_{2i} (y_L)_{i3}\mu_{hSS}|^2}{(4\pi)^4}
\left[(m_h^2 - m_\mu^2 - m_\tau^2)(m_\mu^2 F_L^2 + m_\tau^2 F_R^2)  -4 m_\mu^2 m_\tau^2 F_L F_R \right] ,\\
&F_L=\int\frac{\delta(x+y+z-1) y^2 dxdydz}{(z^2-z) m_\mu^2 + (x^2-y) m_\tau^2 -xz(m_h^2 - m_\mu^2 - m_\tau^2) + x M_{\Psi^{m+1}}^2+(y+z) m_{S^m}^2},\\
&F_R=\int\frac{\delta(x+y+z-1)  z^2 dxdydz}{(z^2-z) m_\mu^2 + (x^2-y) m_\tau^2 -xz(m_h^2 - m_\mu^2 - m_\tau^2) + x M_{\Psi^{m+1}}^2+(y+z) m_{S^m}^2},
\label{eq:hmutau}
\end{align}
where $\mu_{hSS}\equiv \lambda_{\Phi S} v/2$ is the strength of the trilinear $h S^{\pm m}S^{\mp m}$ interaction.
Then the branching ratio reads
\begin{align}
 {\rm BR}(h\to \mu\tau) \approx \frac{ \Gamma(h\to \mu\tau) }{ \Gamma(h\to \mu\tau) + \Gamma(h) },
\end{align}
where $ \Gamma(h) \approx4.2\times 10^{-3}$ GeV is the total decay width of the SM Higgs boson at 125.5 GeV.

\section{ The case $m=0$}
Let us first consider briefly the case $m\neq 0$ before we discuss more constrained model with $m=0$. 
DM candidate does not exist in this case.
Nonzero $m$, however, can enhance the muon $g-2$ as well as the LFVs.
Thus one can obtain the sizable value of muon $g-2$ which can as large as ${\cal O}(10^{-9})$, while one can assume that
the Yukawa coupling matrix ($y_L$) is diagonal {or at least one of $m_{S^m}, m_{\Psi^{m+1}}$'s are very
 large} to evade the constraints of LFVs. 
{
The model has all the ingredients to generate {Majorana} neutrino mass matrix via radiative seesaw mechanism, which, however, is
not straightforward due to the Dirac nature of $L'$~\cite{scoto}.
}
In this sense, radiative neutrino models with (at least) two-loop diagrams are favored for nonzero $m$. 
Another difficulty is decays of $S^{m}$, which is charged scalar for $m \neq 0$, into the SM fields is not possible.  
Some more additional fields need to be introduced in order to evade this problem for each $m$, and the detailed
phenomenology depends on the implemented models. Thus we do not discuss the case of nonzero $m$ further.

We will focus on the special case of $m=0$ because it includes a DM candidate $S^m\equiv S^0$ in the boson sector,
which can possibly solve the above mentioned problems of $m \neq 0$ case.
Notice here that the neutral component of the SU(2)$_L$-doublet $L'$ fermion cannot be DM due to the interaction with the SM neutral gauge
boson $Z$ that is ruled out by the direct detection search. 

We redefine the exotic fields as $(S^0\equiv) S=(S_R+iS_I)/{\sqrt2}$, $L'\equiv [N,E]^T$.
The Lagrangian in (\ref{Eq:lag-flavor}) can be rewritten as
\begin{align}
-\mathcal{L}_{Y}
&=
(y_{\ell})_{ij} \bar L_{Li} \Phi e_{Rj}  + (y_L)_{ij} \bar L_{L_i} L'_{R_j} S
 + { (M_{L})_{ij}} \bar L'_{Li} L'_{Rj} + {\rm h.c.}, \\
{\cal V}&=
m^2_{\Phi} \Phi^\dag\Phi + (m^2_{S_1} S^{2} +{\rm h.c.}) + m^2_{S_2} |S|^2 
+ \lambda_{\Phi} |\Phi^\dag\Phi|^2\nn\\  
&+\sum_{i=0}^4\left[\lambda_i S_i (S^*)^{4-i}+{\rm h.c.}\right]
+ (\lambda_{\Phi S_1} |\Phi|^2 S^2 + {\rm h.c.}) +  \lambda_{\Phi S_2} |\Phi|^2 |S|^2,
\label{Eq:lag-flavor2}
\end{align}
where the corresponding trilinear coupling $\mu_{hS^{\pm m}S^{\mp m}}$ appearing on Eq.~(\ref{eq:hmutau}) is rewritten by $\mu_{hS_RS_R}\equiv (\lambda_{\Phi S_1}+\frac{\lambda_{\Phi S_2}}{2})v$ and $\mu_{hS_IS_I}\equiv (-\lambda_{\Phi S_1}+\frac{\lambda_{\Phi S_2}}{2})v$.
The formulae for LFVs, muon $g-2$, and the excess of $h\to \mu\tau$ are obtained simply by  putting $m=0$ in
Eqs.~(\ref{eq:lfvs}),~(\ref{eq:damu}), and~(\ref{eq:hmutau}). Now we discuss the property of a DM candidate $S_{R/I}$ in the next subsection.

\subsection {Dark Matter Candidate}
Before the analysis of the DM candidate let us make some assumptions for simplicity as follows: 
$X(=S)\equiv S_R \, {\rm or} \,S_I$ ($M_X\equiv m_{S_R}\approx m_{S_I}$), $\mu_{hSS}\equiv \mu_{hS_RS_R}\approx
\mu_{hS_IS_I} \approx \frac{\lambda_{\Phi S_2}v}{2}$, therefore $\lambda_{\Phi S_1}\ll \lambda_{\Phi S_2}$. 

\begin{figure}[tb]
\begin{center}
\includegraphics[width=0.6\textwidth]{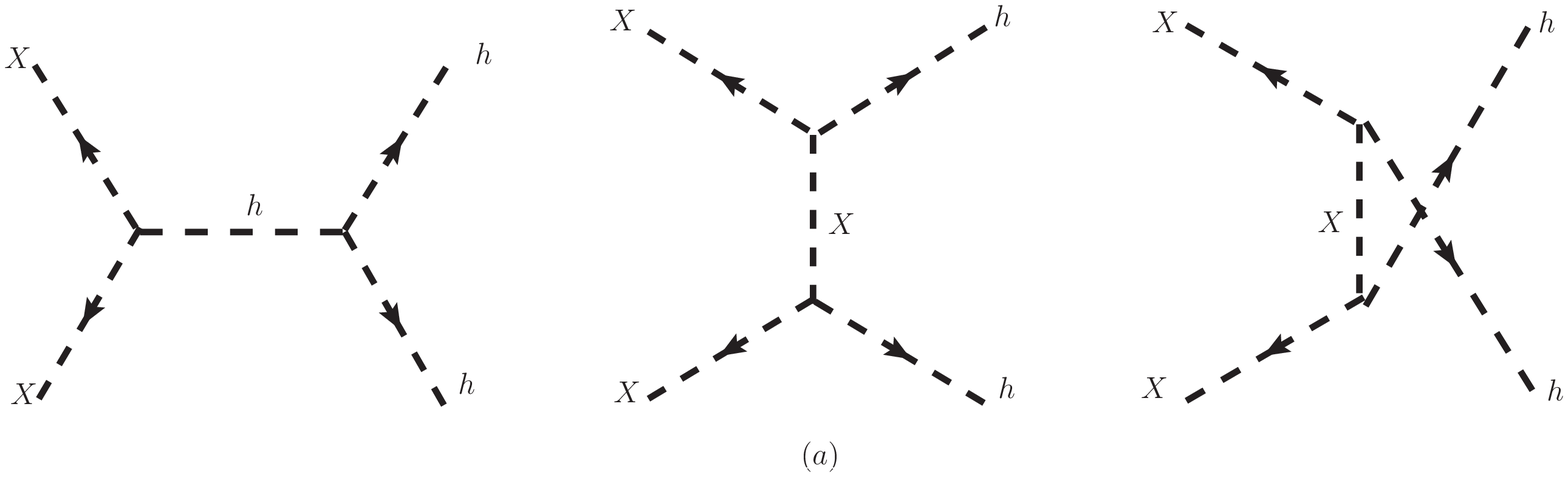}
\includegraphics[width=0.25\textwidth]{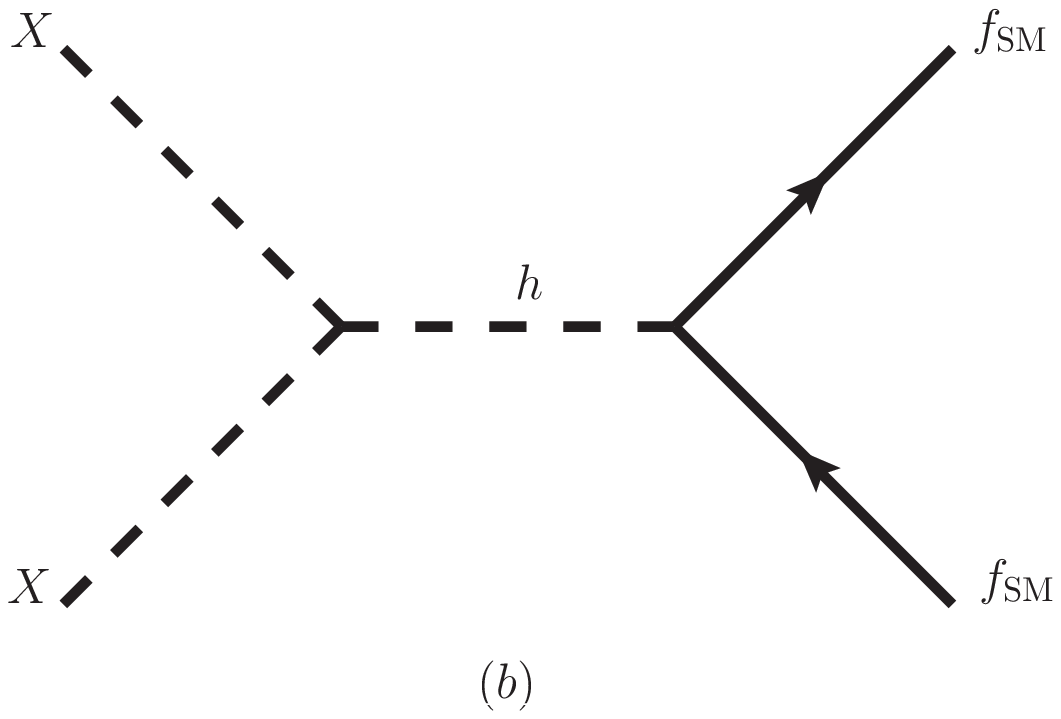}
\includegraphics[width=0.25\textwidth]{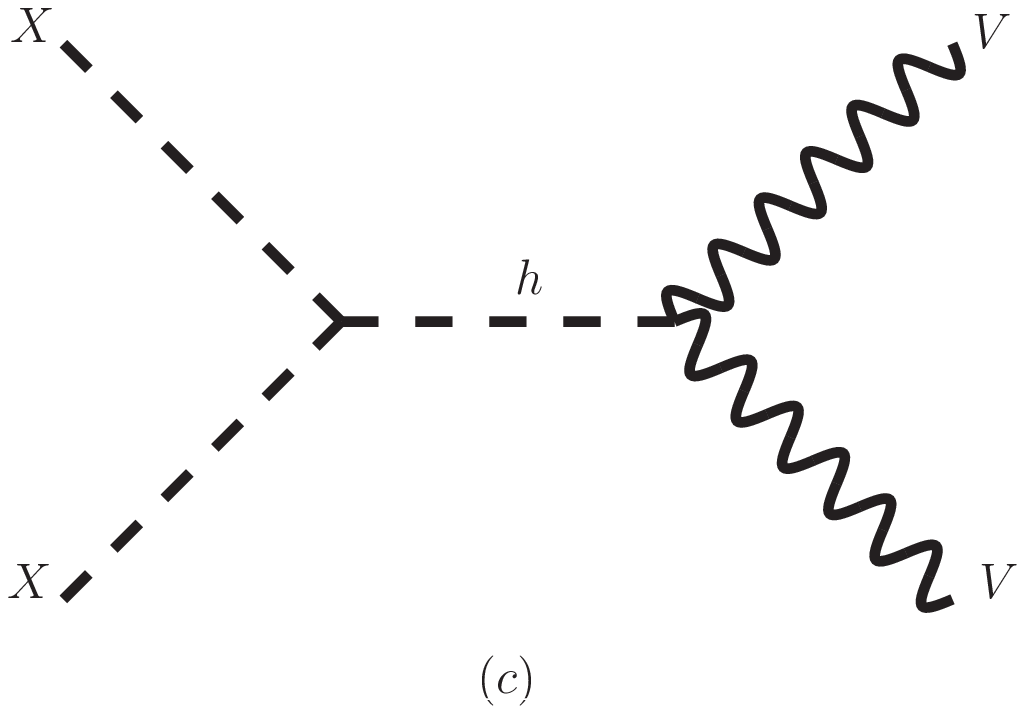}
\includegraphics[width=0.4\textwidth]{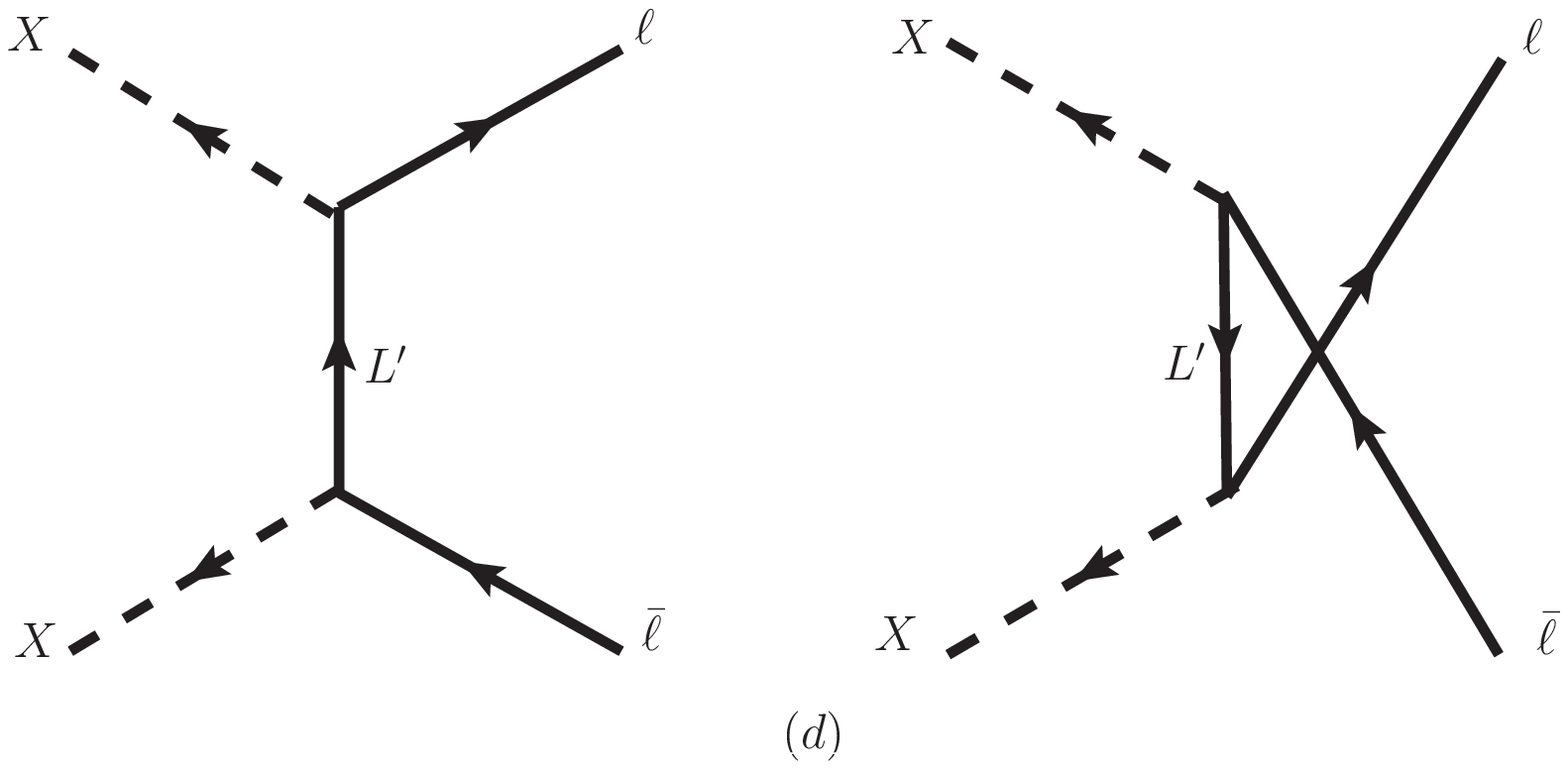}
\caption{Feynman diagrams for the DM annihilations.}
\label{fig:ann}
\end{center}
\end{figure}

{\it Relic density}: 
{The thermal averaged annihilation cross section comes from the processes $2X\to 2h$, 
$2X \to f_{SM} \bar f_{SM}$, $2X \to V V^{(*)}$,
and $L'$-exchanging $2X\to \ell\bar\ell (\nu_L\bar\nu_L)$~\cite{Griest:1990kh, Edsjo:1997bg}, $f_{SM}$ and $V$ being the SM fermions and
gauge bosons, respectively. The Feynman diagrams are shown in Fig.~\ref{fig:ann}.} It can be calculated as
\begin{align}
&\sigma v_{\rm rel}\approx \sum_{f=h,f_{SM},\ell,V}\int_0^{\pi} \sin\theta d\theta \frac{ |\bar M|^2}{16\pi s}\sqrt{1-\frac{4 m_f^2}{s}},
\end{align}
where
\begin{align}
& |\bar M|^2\approx  |\bar M(2X\to 2h)|^2+ \sum_{f_{SM}=(t,b)}|\bar M(2X\to f_{SM} \bar f_{SM})|^2\nn\\
& \hspace{3cm} +  \sum_{\ell=(\ell,\nu_L)}|\bar M(2X\to \ell\bar \ell)|^2+
  \sum_{V=(Z,W^\pm)} |\bar M(2X\to V V^*)|^2,\\ 
& |\bar M(2X\to 2h)|^2
\approx \lambda_{\Phi S_2}^2 
\left|
1+\frac{3v^2 \lambda_\Phi}{2(s-m_h^2)} + \frac{v^2}{4}
\left[\frac{1}{t-M_X^2}+\frac{1}{u-M_X^2}\right] 
\right|^2,
\\
& |\bar M(2X\to f_{SM}\bar f_{SM})|^2
\approx \frac{48 \mu_{hSS}^2 m_{f_{SM}}^2}{(s-m_h^2)^2 v^2}\left(\frac s2-2 m_{f_{SM}}^2\right), 
\label{eq:XX2ff}\\
& |\bar M(2X\to VV^*)|^2
\approx \frac{4 \lambda_{\Phi S_2} \mu_{hSS}^2 m_V^4}{(s-m_h^2)^2 v^2}\left(2 + \frac{(s/2- m_{V}^2)^2}{m_V^4}\right), \\
%
%
%
& |\bar M(2X\to \ell\bar \ell)|^2\approx
8 \sum_{a,b}^{2-3}\sum_{i=1-3}|(y_L)_{i,b}|^2 |(y_L)_{i,a}|^2\times \nn\\
&
\left[
4\left(\frac{p_1\cdot k_1}{t}+\frac{p_2\cdot k_1}{u} \right) \left(\frac{p_1\cdot k_2}{t}+\frac{p_2\cdot k_2}{u} \right)
-s M_X^2 \left(\frac{1}{t^2}+\frac{1}{u^2}\right) -2 s \left(\frac{p_1\cdot p_2}{tu}\right)
\right],
\end{align}
where $s,t,u$ are the Mandelstam variables; $p_1,p_2 (k_1,k_2)$ are four-momenta of the initial (final) states; 
$\lambda_{\Phi S_2} = {2\mu_{hSS}}/{v}$;  among all the SM fermions in $f_{SM}$  heavy quarks such as top quark or bottom quark dominate; 
$V(=Z,W^\pm)$ is the SM vector gauge bosons. 
We neglect the masses of the SM leptons ($\nu_L,\ell$) in the final states.
Notice here that the mode  $ 2X\to \ell\bar \ell$ is $d-$wave dominant. 
To include its effect we retain terms up to the  $v_{\rm rel}^4$ in $v_{\rm rel}$ expansion for all the modes.
Then the relic density of DM is finally obtained from
\begin{align}
\Omega h^2\approx \frac{1.07\times10^9}
{g^{1/2}_* M_{\rm pl}[{\rm GeV}] \int_{x_f}^\infty \left(\frac{a_{\rm eff}}{x^2}+6\frac{b_{\rm eff}}{x^3} +60\frac{d_{\rm eff}}{x^4} \right)},
\end{align}
where $g_*\approx 100$ is the total number of effective relativistic degrees of freedom at the time of freeze-out,
$M_{\rm pl}=1.22\times 10^{19}[{\rm GeV}] $ is the Planck mass, $x_f\approx25$, and $ a_{\rm eff}$, $ b_{\rm eff}$ and $d_{\rm eff}$ are coefficients
in the  $v_{\rm rel}^2$ expansion of the annihilation cross section:
\begin{align}
\sigma v_{\rm rel}\approx a_{\rm eff}+ b_{\rm eff} v^2_{\rm rel} + d_{\rm eff} v^4_{\rm rel}.
\end{align}
The observed relic density reported by Planck suggest that $\Omega h^2\approx 0.12$~\cite{Ade:2013zuv}.
{In terms of the model parameters the expansion coefficients are
\begin{align}
a_{\rm eff} &\approx
\frac{3\mu_{hSS}^2 \sum_{f=b,t} m_f^2(M_X^2-m_f^2) }{2\pi M_X^2 v^2(m_h^2-4M_X^2)^2}
\sqrt{1-\frac{m_f^2}{M_X^2}}\nn\\
&+\frac{\lambda_{\Phi S}^2 }{256\pi M_X^2}
\left|2+v^2\left( \frac{1}{m_h^2-2M_X^2} -\frac{3\lambda_\Phi}{m_h^2-4M_X^2} \right)\right|^2
\sqrt{1-\frac{m_h^2}{M_X^2}}\nn\\
 &+
\frac{3\mu_{hS S }^2 \sum_{V=W,Z}}
{16\pi M_X^2 v^2(m_h^2-4M_X^2)^2}
\left(2m_V^4 +(2M_X^2-m_V^2)^2\right) 
\sqrt{1-\frac{m_V^2}{M_X^2}},
\end{align}
where we would not show the explicit forms of $b_{\rm eff}$ and $d_{\rm eff}$, because they are too complicated. 
}

{\it Direct detection}: 
The DM-nucleon scattering is induced by the SM Higgs exchanging process in our model, which is calculated in
non-relativistic limit. 
The dominant tree-level diagram is obtained by crossing Fig.~\ref{fig:ann} (b) which gives (\ref{eq:XX2ff}).
However, the leptons in $f_{SM}$ in the crossed diagram does not contribute to the direct detection process because there
is no valence leptons inside nucleons.  Although heavy quark contributions to the parton distribution function of
nucleon are suppressed, they can make contribution via Higgs-gluon-gluon triangle diagram.
Here we estimate the DM-nucleon scattering cross section following Ref.~\cite{Cline:2013gha}.
 Firstly we obtain the following effective Lagrangian by integrating out $h$ for non-relativistic momentum transfer,
\begin{align}
\mathcal{L}_{\text{eff}}
=
\sum_{{q}} \frac{C_{hSS} m_q}{m_h^2} X^2 \bar{q} q,
\end{align}
where $q$ and $m_q$ represent the corresponding quark fields and the quark masses respectively, the sum is over all
quark flavors, and we neglected higher dimensional operators.
The coefficient $C_{hSS}$ determines the effective interaction between the quarks and $X$.
The corresponding value in our model is
\begin{align}
C_{hSS} = \frac{\mu_{hSS}}{v}.
\end{align}
%
Then the effective $X$-nucleon ($N$) interaction can be written down by 
\begin{equation}
\mathcal{L}^{N}_{\text{eff}} = \frac{ f_N C_{hSS} m_N}{m_h^2} X^2 \bar N N
\end{equation}
where the effective coupling constant $f_N$ is given by
\begin{equation}
f_N = \sum_{q} f_q^N = \sum_{q} \frac{m_q}{m_N}  \langle N | \bar q q | N \rangle.
\end{equation}
Note that the quark mass $m_q$ is absorbed into the definition of quark mass fraction
\begin{align}
f_q^N \equiv  {m_q \over m_N} \langle N | \bar q q | N \rangle.
\end{align}
The heavy quark contributions are replaced by the gluon contributions by calculating the triangle diagram
\begin{align}
\sum_{q=c,b,t}  f_q^N = {1 \over m_N} \sum_{q=c,b,t} \langle N | \left(-{ \alpha_s\over 12 \pi} m_q G^a_{\mu\nu}
  G^{a\mu\nu}\right)| N \rangle.
\label{eq:f_Q}
\end{align}
From the scale anomaly, the trace of the stress energy tensor is written as~\cite{Shifman:1978zn}
\begin{align}
\theta^\mu_\mu =m_N \bar{N} N = \sum_q m_q \bar{q} q - {7 \alpha_s \over 8 \pi} G^a_{\mu\nu} G^{a\mu\nu}.
\label{eq:stressE}
\end{align}
From (\ref{eq:f_Q}) and (\ref{eq:stressE}) we finally obtain
\begin{align}
\sum_{q=c,b,t} f_q^N = \frac{2}{9} \left( 1 - \sum_{q = u,d,s} f_q^N \right),
\end{align}
which results in
\begin{align}
f_N = \frac29+\frac{7}{9}\sum_{q=u,d,s}f_{q}^N. 
\end{align}
%
Here we use the DM-neutron ($n$) scattering cross section to consider constraints from direct detection where that of DM-proton case is almost same for Higgs portal interaction.
Then the spin independent scattering cross section of the {$X$} with neutron through the SM Higgs($\phi$) portal
process is obtained to be~\cite{Cline:2013gha}
\begin{align}
\sigma_{\rm SI}(Xn \to Xn) \times \left( \frac{\rho_{X}}{\rho_{DM}} \right) &
=\frac{1}{ \pi}\frac{\mu^2_{nX}}{M_X^2} \frac{ m_n^2 C_{hSS}^2 f_n^2}{m_h^4} \times \left( \frac{\rho_{X}}{\rho_{DM}} \right)
\simeq \frac{\mu_{h S S}^2 f_{n}^2}{\pi v^2} \frac{ m_n^4}{m_h^4 M_X^2}
\left(\frac{\Omega h^2}{0.12}\right) \nn\\
&\approx 5.29\times 10^{-43}\left(\frac{\mu_{hSS} }{M_X}\right)^2
\left(\frac{\Omega h^2}{0.12}\right) \ [{\rm cm}^2],\label{eq:dd}
\end{align}
where $m_n$ is the neutron mass, we approximated $\mu_{nX} = m_n M_X/(m_n+M_X) \simeq m_n$, $\rho_X$ and $\rho_{DM}$ are
current density of $X$ and
total density of DM,  and  $f_{n} \approx 0.287$(with $f_{u}^n = 0.0110, \, f_{d}^n = 0.0273, \,
f_{s}^n = 0.0447$) represents the sum of the contributions of partons to the mass fraction of
neutron~~\cite{Belanger:2013oya}. 
The scattering cross section imposes a strong constraint on the parameter space relevant to the DM.
The constraint from the  LUX experiment is the strongest at present with 
$\sigma_{\rm SI} \times \left( \frac{\rho_{X}}{\rho_{DM}} \right)$ less than ${\cal O}$(10$^{-45}$) cm$^2$ for DM mass about ${\cal O}$(10) GeV~\cite{Akerib:2013tjd}.
{Notice here that the experimental bound on the direct detection is obtained by assuming that one of the DM components occupies all of the DM components~({\it i.e.}, $\Omega h^2=0.12$) in the current universe.  Otherwise we should multiply the factor $\left( \frac{\rho_{X}}{\rho_{DM}} \right) = \left(\frac{\Omega h^2}{0.12}\right)$ for the scattering cross section as can be seen in Eq.~(\ref{eq:dd}).
 It suggests that the upper bound from direct detection is relaxed when our $X$ is subcomponent of DM, because $1 > \left(\frac{\Omega h^2}{0.12}\right)$. }


{\it Numerical analysis}: 
\begin{figure}[tb]
\begin{center}
\includegraphics[scale=0.6]{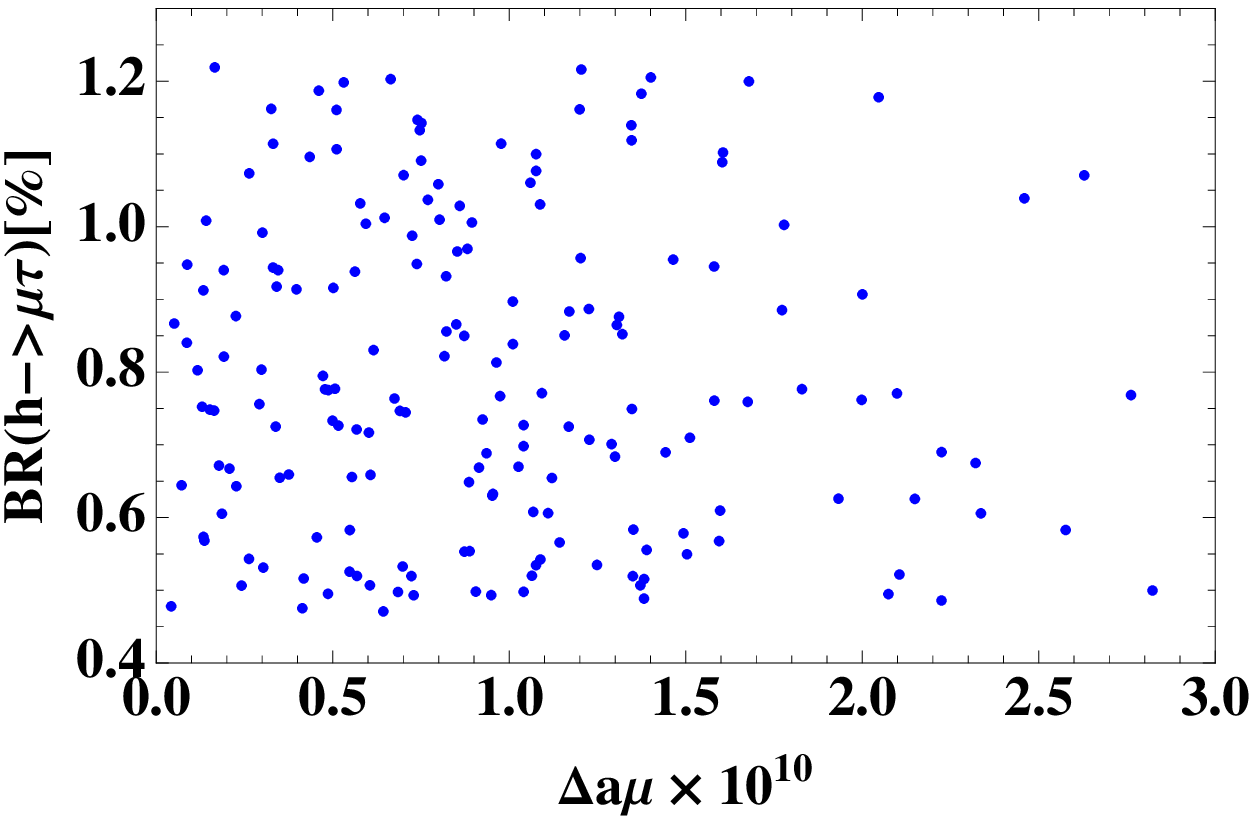}
\includegraphics[scale=0.6]{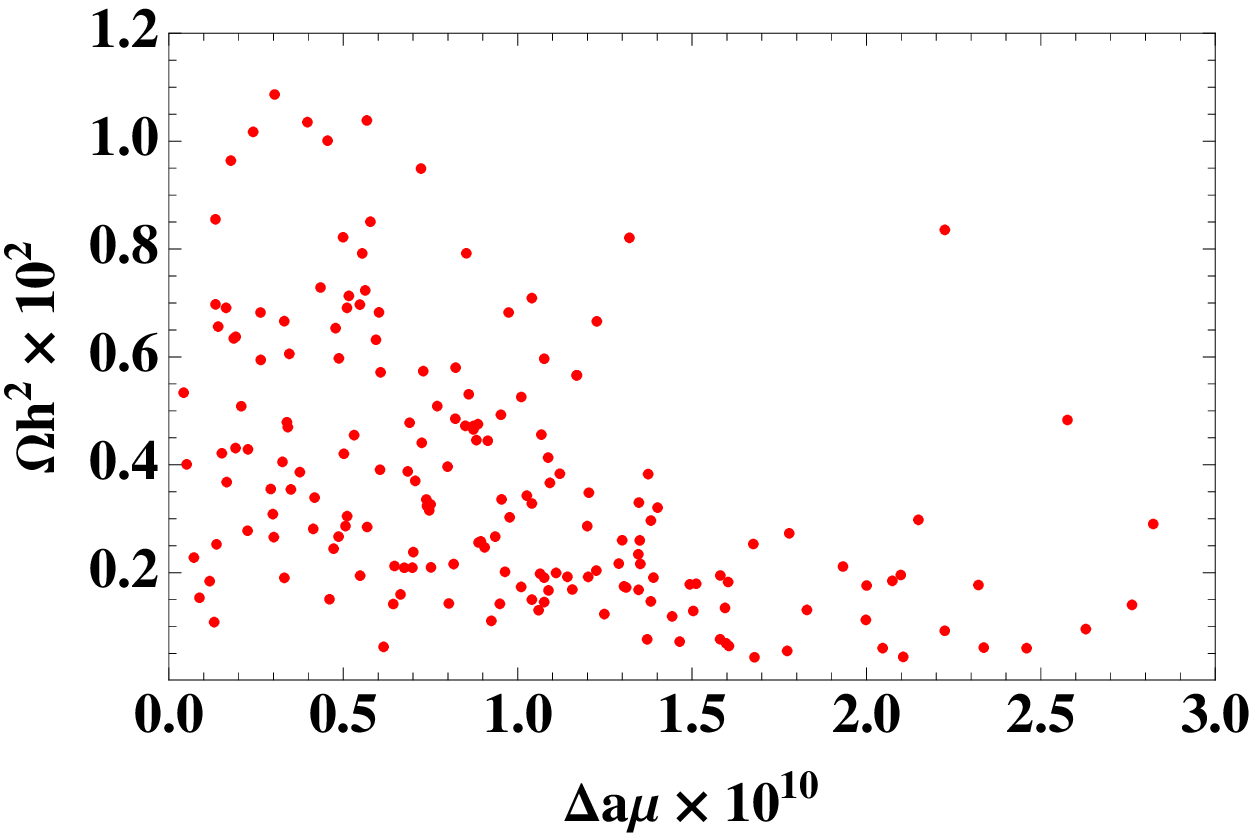}
\includegraphics[scale=0.6]{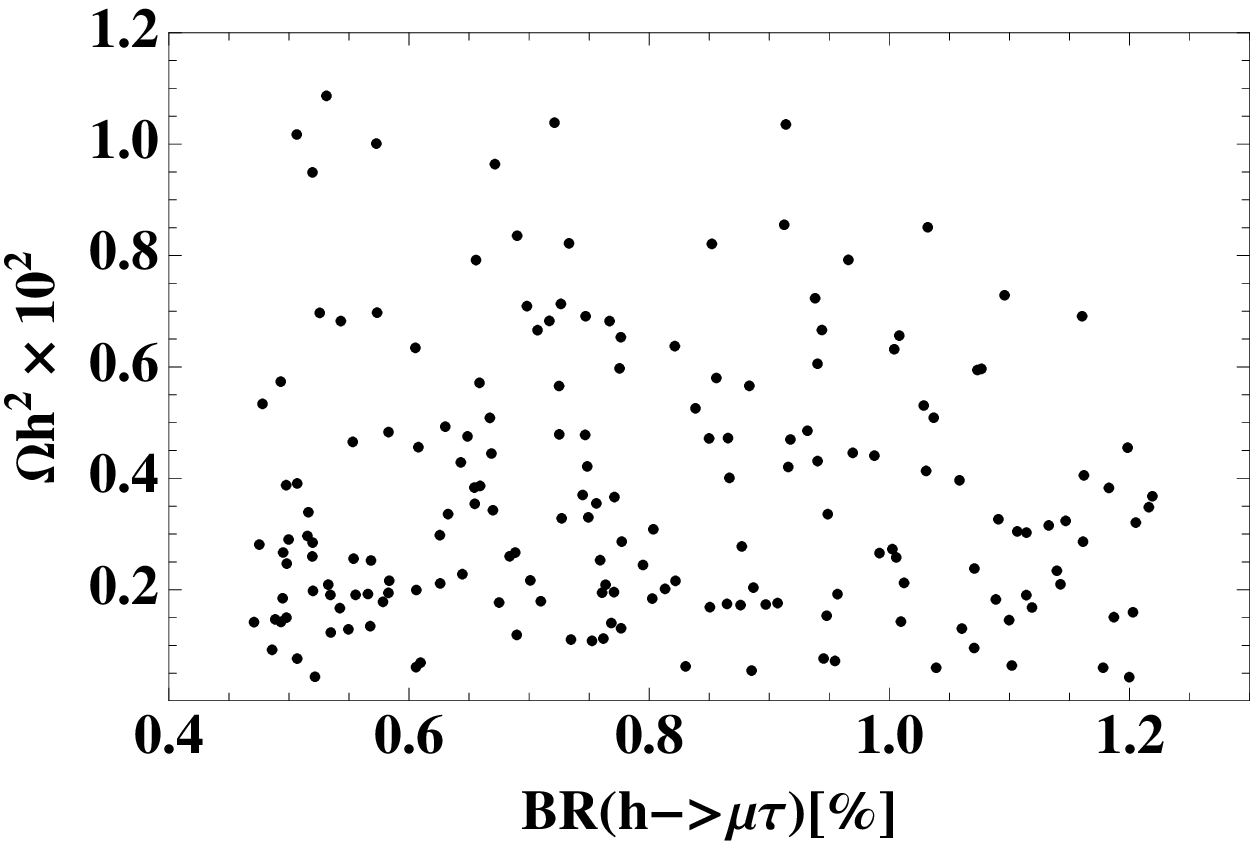}
\caption{Numerical results: 
Each of the top-left, top-right, and bottom figure represents the scattering points in terms of (muon $g-2$ and the branching ratio of $h\to\mu\tau$), (relic density and the muon $g-2$), and (the branching ratio of $h\to\mu\tau$ and relic density).
 }
\label{num:min}
\end{center}
\end{figure}
Now that all of the analytical formulae are derived, we perform numerical analysis and explore the allowed region.
We scan the parameters in the ranges:
\begin{align}
& M_X \in [100\,\text{GeV}, 500\,\text{GeV}],\quad \mu_{hSS} \in [50\,\text{GeV}, 500\,\text{GeV}],\quad
M_L (=M_{E_i}=M_{N_i}) \in [M_X,\ 1\,\text{TeV}],\nn\\
& (y_L)_{\ell,m} \in [-0.01,0.01],\ (\ell,m)=((1,1),(2,1),(3,1)), \quad (y_L)_{i,j} \in [-\sqrt{4\pi},4\pi],\ (i,j)\neq (\ell,m), 
	\label{range_scanning}
\end{align}
where $M_X = M_S$ denotes DM mass.
{The ranges are chosen to satisfy perturbativity of $\lambda_{\Phi S}$, the bound from the charged lepton
  flavor violation, and also electroweak scale new particles are assumed. The result does not change much even if we enlarge
  the ranges.}
{Here we assume Yukawa couplings $(y_L)_{ij}$ are small when it has index corresponding to electron in order to satisfy constraint from $\mu \to e \gamma$. }

We scanned the above regions of parameters randomly to obtain the allowed range of $h\to \mu\tau$ branching ratio and muon $g-2$, imposing the
constraint from dark matter relic density and direct detection experiments.
The results are shown in Fig.~\ref{num:min}. As we can see in the figure, we can accommodate the excess observed by CMS.
 In this case, the maximum value of the muon $g-2$ is around $(3\times 10^{-10})$, which is smaller than the current
 discrepancy  ${\cal O}(10^{-9})$~\cite{bennett}.
The relic density is  ${\cal O}(0.001-0.01)$, which is also smaller than the current measurement 0.12 as can be seen in
Fig.~\ref{num:min}. It is mainly due to  the direct detection bound. In order to obtain enough excess of $h\to \mu\tau$, one has to increase the value of trilinear coupling $\mu_{hSS}$.  
On the other hand,  the direct detection bound suggests $\frac{\mu_{hSS}}{M_X}\lesssim 0.04$, if  $\sigma_{SI}
\lesssim 10^{-45}$ cm$^2$. To evade the constraint from LUX the DM mass scale should be above 100 TeV, which conflicts with  
relic density and $h\to\mu\tau$ excess.
This leads to the conclusion that $S$ cannot be main source of the relic DM.

\begin{figure}[tb]
\begin{center}
\includegraphics[scale=0.6]{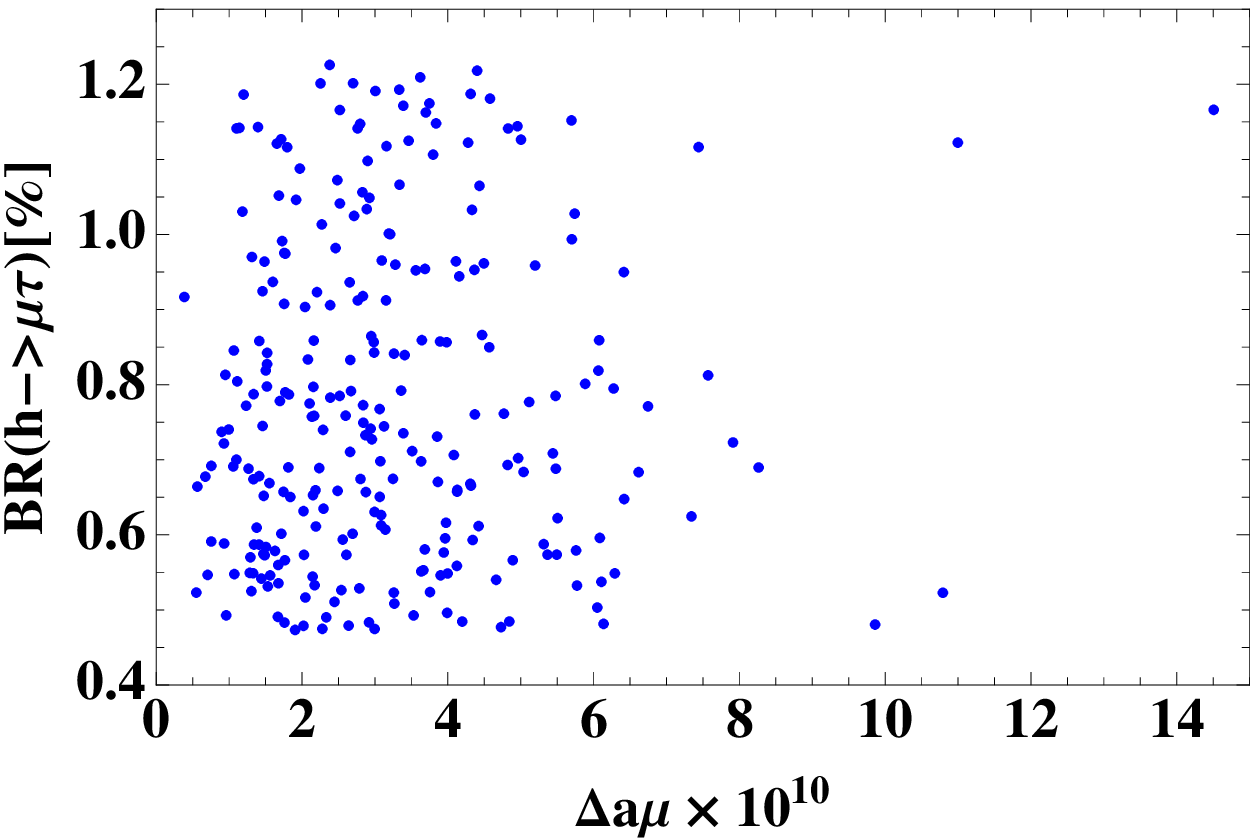}
\includegraphics[scale=0.6]{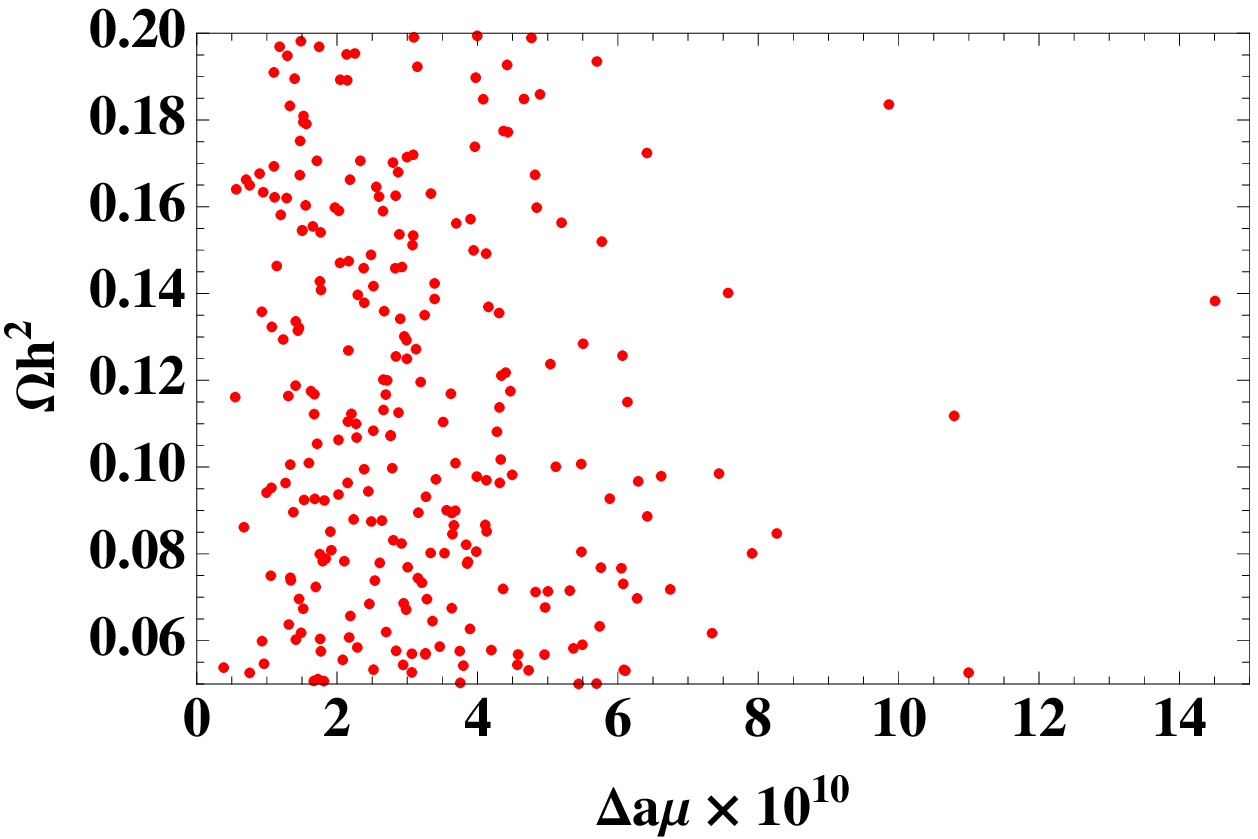}
\includegraphics[scale=0.6]{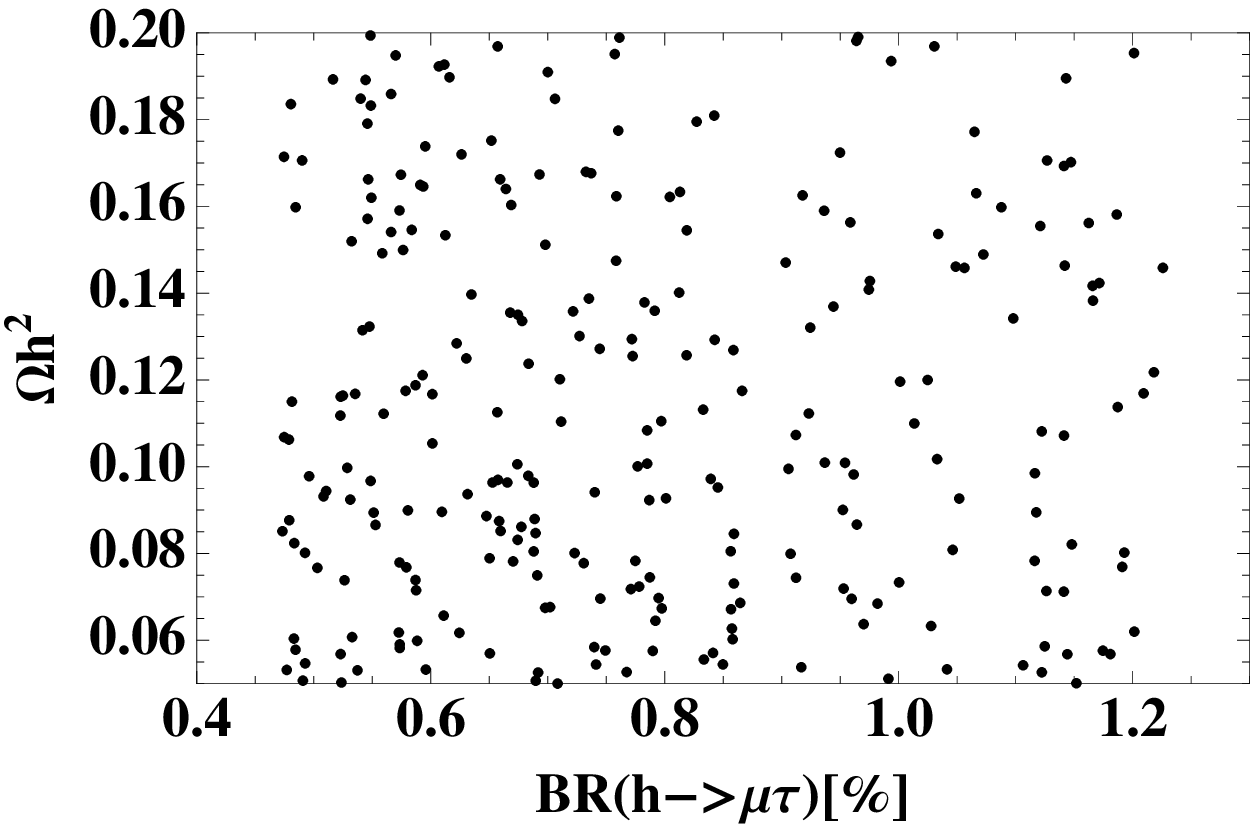}
\caption{Numerical results: 
Each of the top-left, top-right, and bottom figure represents the scattering points in terms of (muon $g-2$ and the branching ratio of $h\to\mu\tau$), (relic density and the muon $g-2$), and (the branching ratio of $h\to\mu\tau$ and relic density). }
\label{num:ex}
\end{center}
\end{figure}

To explain DM relic abundance we extend the  model as minimally as possible.
One of the minimal extension to solve this issue is to introduce another gauge singlet boson having the same charge with
$S$, which we denote by $S_2$. 
Then all the terms of Eq.~(\ref{Eq:lag-flavor2}) remain in the same form with only the number of terms doubled.
For our convenience let us rename two singlet bosons  $(S_1,S_2)$. Then the Lagrangian is simply obtained by replacing,
e.g., $y_L\to y_{L}^\alpha(\alpha=1-2)$, $\mu_{hSS}\to \mu_{hS_iS_j} (i,j=1-2)$, {\it etc}.
{Explicitly the new terms include
\begin{align}
-\mathcal{L}_{Y}
& \supset
 \sum_{\alpha=1,2}\left((y_L^\alpha)_{ij} \bar L_{L_i} L'_{R_j} S_\alpha 
 - \sum_{\beta=1,2} \lambda_{\Phi S_\alpha S_\beta} |\Phi|^2 S_\alpha S_\beta
 + {\rm h.c.}\right), 
\label{Eq:lag-flavor}
\end{align}
where $\mu_{hS_\alpha S_\beta}\equiv \lambda_{\Phi S_\alpha S_\beta }v/2$ and we neglect $ \mu_{hS_1S_2}$ for simplicity.
}
We assume $M_{S_2} > M_{S_1}$ so that $S_1$ still remains as a DM candidate.
In this case, $\mu_{hS_2S_2}$  can play a crucial role in generating the excess $h\to \mu\tau$, 
while it need not contribute to the interaction of the  direct detection searches.
We take the same regions given Eq.~(\ref{range_scanning}) as our new input parameters except the following, 
\begin{align}
& \quad M_{S_2} \in [\frac{11}{10}\,\text{GeV}, 1 \,\text{TeV}], \quad  \mu_{hS_1S_1} \in [0.01\,\text{GeV}, 0.1\,\text{GeV}],
\label{range_scanning-2}
\end{align}
and $M_X = M_{S_1}$ in this case.
We show the results in Fig.~\ref{num:ex}, in which relic density is within the current observational value.
The maximum value of the muon $g-2$ is around $1.5\times 10^{-9}$, which can explain the discrepancy $\Delta a_\mu =
(26.1 \pm 8.0) \times 10^{-10}$~\cite{discrepancy1} at the 2$\sigma$ level.
To obtain $O(10^{-9})$ muon $g-2$, $M_X \in [100 \, \text {GeV}, 170 \, \text {GeV}]$ and $\mu_{hS_2S_2} \in [20\,\text{GeV}, 150\,\text{GeV}]$ are
preferred. 
For simplicity we assume the mass difference ratio $(M_{S_2} - M_X)/M_X \gtrsim 10 \%$ to evade the coannihilation regime.
The trilinear coupling $\mu_{hS_1S_1}$ can be decreased by three orders magnitude below the original value of one $S$
model to satisfy the direct detection experiments without affecting other observables,
$h \to \mu\tau$, $(g-2)_\mu$, DM relic density. In this case the DM relic density is achieved dominantly by $2X \to
\ell\bar{\ell}$ channel.
{
Here we provide the typical parameter set as follows:
\begin{align}
& M_X \approx 146\ \text{GeV},\quad 
M_L (=M_{E_i}=M_{N_i}) \approx  (663,980,460) [\text{TeV}],\nn\\
&
M_{S_2} \approx332 [\text{TeV}] , \quad \mu_{hS_1S_1} \approx 0.079 [{\rm GeV}],
\quad \mu_{hS_2S_2} \approx 23 [{\rm GeV}], \nn\\
& (y_L)_{\ell,m} \approx 
 \left[
\begin{array}{ccc}
-0.0076 & -1.0 & 0.16 \\
-0.0076 &	-0.83 & -2.8 \\
-0.0063 & 0.38 & -2.4 \\
\end{array}\right],\
  (y_L)_{i,j} \approx \left[
 \begin{array}{ccc}
-0.0060 & -0.89 & -3.0 \\
-0.0062 &	-0.25 & -3.3 \\
0.0 & 2.5 & -0.38 \\
\end{array}\right],
 \end{align}
then we can obtain the following observables:
\begin{align}
&(g-2)_\mu\approx 1.8\times 10^{-10},\quad \Omega h^2\approx 0.12,
\quad {\rm BR}(h\to\mu\tau)\approx0.48\%,\quad
 \sigma_{\rm SI}\approx 4.4\approx 10^{-50}\ [{\rm cm}^2],\nn\\
&{\rm BR}(\mu\to e\gamma)\approx 2.5\times10^{-13},
\quad
{\rm BR}(\tau\to e\gamma)\approx 8.6\times10^{-12},
\quad
{\rm BR}(\mu\to\mu\gamma)\approx 1.3\times10^{-8}.
\quad
\end{align}
}

\begin{figure}[tb]
\begin{center}
\includegraphics[width=70mm]{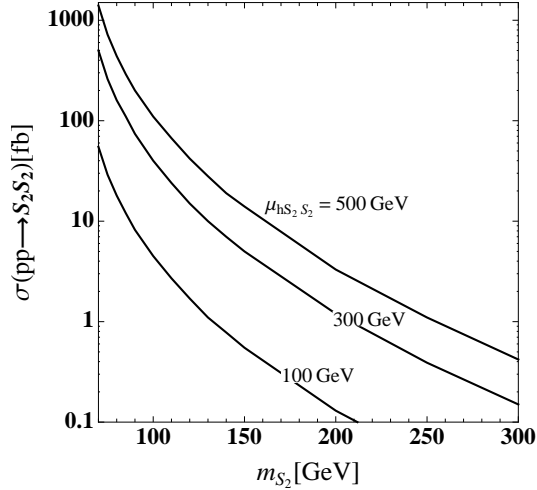}
\caption{The production cross section for $pp \to S_2 S_2$ as a function of $m_{S_2}$ with collision energy of $\sqrt{s}=14$ TeV.
 }
\label{fig:XS}
\end{center}
\end{figure}
{\it Collider phenomenology of our scenario:}
Now we discuss the signature of our scenario at the LHC.
Here we focus on the interaction $\mu_{hS_2S_2} h |S_2|^2$ to produce $S_2$ via gluon fusion, $gg \to h \to S_2 S_2$, since the coupling constant of the interaction $h S_2 S_2$ is required to be large as $O(100)$ GeV in obtaining sizable $h \to \mu \tau$ branching ratio. 
The produced $S_2$ mainly decays into $S_1 h$ through the interaction like $v h S_1 S_2$ in scalar potential where we assume $L'$ is heavier than $S_2$ for simplicity.
It suggests that the $S_2$ can be measured at the LHC where the signature will be two SM Higgs boson with missing transverse energy.
{Then the production cross section is numerically estimated with CalcHEP~\cite{Belyaev:2012qa} using {\tt CTEQ6L} PDF~\cite{Nadolsky:2008zw} by implementing relevant interactions in the code.}
In Fig.~\ref{fig:XS}, we show the production cross section of $pp \to S_2 S_2$ as a function of $S_2$ mass adopting some values of $\mu_{hS_2S_2}$ and collision energy of $\sqrt{s}=14$ TeV.
{Here the cross section is at the leading order and it will be larger when we consider K-factor. }
We find that the cross section can be sizable when $\mu_{hS_2S_2}$ is large and $m_{S_2}$ is around 100 GeV. 
{Note that the cross section becomes significantly large when the $m_{S_2}$ close to $m_h/2$ due to resonant enhancement
  since the process is SM Higgs boson exchanging s-channel, although the resonant point is below our parameter region.}
Thus some parameter space of our scenario can be tested by exploring $h h \ET$ signal at the LHC. 
{In Table~\ref{tab:event}, we also show the number of expected events at the LHC 14 TeV for several values of $m_{S_2}$ and $\mu_{h S_2 S_2}$ with luminosity of 100 fb$^{-1}$ as a reference.}
Moreover the study of exotic lepton production will be also interesting.
The detailed simulation study is beyond the scope of this paper and it is left as future study.
 \begin{widetext}
\begin{center} 
\begin{table}[ht]
\begin{tabular}{|c| c c c |}\hline
& $m_{S_2}$ = 100 GeV & $m_{S_2}$ = 200 GeV & $m_{S_2}$ = 300 GeV  \\ \hline
$\mu_{h S_2 S_2}=$ 100 GeV & 4.5$\times 10^2$  & 13.  & 1.7  \\ 
$\mu_{h S_2 S_2}=$ 200 GeV & 4.0$\times 10^3$  & 1.2$\times 10^2$   & 15. \\ 
$\mu_{h S_2 S_2}=$ 300 GeV & 1.1$\times 10^4$  &  3.3$\times 10^2$  & 42.     \\ \hline
\end{tabular}
\caption{Number of expected $hh \ET$ events at the LHC 14 TeV for several values of $m_{S_2}$ and $\mu_{h S_2 S_2}$ with luminosity of 100 fb$^{-1}$.}
\label{tab:event}
\end{table}
\end{center}
\end{widetext}
%


\section{ Conclusions and discussions}
We studied a possibility to explain the excess of $h\to \mu\tau$ and muon $g-2$ in a model with a dark matter candidate.
At first, we provided a simple set up with generic hypercharge assignments, in which we formulated the lepton flavor
violations, muon $g-2$, and the  branching ratio of $h\to\mu\tau$. 
Then we moved on to the specific case where single DM candidate can  be included. We found the sizable excess of
$h\to\mu\tau$ has been obtained. However the relic density and muon $g-2$ cannot be explained due to the stringent
constraint from the direct detection via Higgs portal. 

We extended the model as minimally as possible so that we can explain the relic density and the muon $g-2$ as well as
$h\to \mu\tau$.
We introduced another gauge singlet boson  $S_2$  having the same quantum numbers with $S_1$, and we solved all the issues.
At the end, we have discussed the signature of our scenario at the LHC, focusing on the interaction $\mu_{hS_2S_2} h |S_2|^2$ to produce $S_2$ via gluon fusion, $gg \to h \to S_2 S_2$.
This is because the coupling constant of the interaction $h S_2 S_2$ is required to be large as $O(100)$ GeV in obtaining sizable $h \to \mu \tau$ branching ratio. 
The produced $S_2$ mainly decays into $S_1 h$ through the coupling $v h S_1 S_2$  in scalar potential. 
It suggests that the $S_2$ can be searched for at the LHC where the signature is two SM Higgs boson with missing
transverse energy. 
We found that the cross section can be sizable when when $m_{S_2}$ is around 100 GeV for $\sqrt{s} =14 $ TeV $p p$ collision.

It is worth to mention  possible application of our model to the other sectors such as neutrinos. 
Since our set up is very simple, several applications to the neutrino sector could be possible. Let us just briefly
comment on  two possibilities. 
First, if we introduce a gauge singlet Majorana fermion with $Z_2$ odd charge, we can explain the neutrino masses  and
mixings at the one-loop level, and the fermion can be a good DM candidate.
Second possibility is to introduce a SU(2) triplet boson  with nonzero VEV.
In this case, the neutrino masses and the mixings are induced through the type-II seesaw mechanism.The neutral component
of the SU(2)$_L$ doublet exotic lepton can be a DM candidate since sizable mass splitting between right-handed and
left handed neutral fermions can be obtained to evade the strong bound from direct detection experiments~\cite{Arina:2012aj}.
However since there is no new source of the muon $g-2$ for both cases, we need some extensions such as we have mainly
discussed in our paper.


\section*{Acknowledgments}
\vspace{0.5cm}
H. O. is sincerely grateful for all the KIAS members, Korean cordial persons, foods, culture, weather, and all the other things.
This work is supported in part by National Research Foundation of Korea (NRF) Research Grant NRF-2015R1A2A1A05001869 (SB).

\end{document}